\pdfoutput=1

\documentclass[sigconf,natbib=true, dvipsnames, svgnames]{acmart}

\AtBeginDocument{%
  \providecommand\BibTeX{{%
    \normalfont B\kern-0.5em{\scshape i\kern-0.25em b}\kern-0.8em\TeX}}}

\usepackage[inline]{enumitem}
\usepackage{pbox}
\usepackage{balance}
\newlist{inlinelist}{enumerate*}{1}
\setlist*[inlinelist,1]{label=\roman*),itemjoin={{, }},itemjoin*={{, and }}}
\usepackage{url}
\usepackage[normalem]{ulem}
\usepackage{multirow}
\usepackage{multicol}
\usepackage{booktabs}
\usepackage{pgfplots}
\usepackage{graphicx}
\usepackage{caption}
\usepackage{subcaption}
\usepackage{multirow}
\usepackage{adjustbox}
\pgfplotsset{compat=1.16}
\usepgfplotslibrary{statistics}
\usepgfplotslibrary{groupplots}

\makeatletter
\@namedef{ver@everyshi.sty}{}
\makeatother

\usepackage{pgfplots, pgfplotstable}
\pgfplotsset{compat=newest}
\usepgfplotslibrary{fillbetween}

\newcommand{\baselineplotc}{Gray}
\newcommand{\separateplotc}{Blue}
\newcommand{\lplotc}{RubineRed}
\newcommand{\plmplotc}{PineGreen}
\newcommand{\smallencoderc}{Magenta}
\newcommand{\noencoderc}{Orchid}
\newcommand{\bmvinc}{CornflowerBlue}
\newcommand{\docTc}{WildStrawberry}
\newcommand{\deepimpacc}{Turquoise}
\newcommand{\unicoilc}{Emerald}

\pgfmathsetmacro{\stdgrad}{30}

\tikzset{every mark/.append style={solid}}
\pgfplotsset{
	grid=both, width=\linewidth, try min ticks=5,
	legend cell align=left, legend style={fill opacity=0.8},
	ylabel near ticks,
    xlabel near ticks,
    every tick label/.append style={font=\footnotesize},
}

\pgfplotsset{
    baselineplot/.style={thick, color=\baselineplotc, mark=x, mark size=3pt},
    separateplot/.style={thick, color=\separateplotc, mark=star, mark size=3pt},
    lplotc/.style={thick, color=\lplotc, mark=triangle*, mark size=3pt},
    plmplotc/.style={thick, color=\plmplotc, mark=pentagon*, mark size=3pt},
    tinybertplot/.style={thick, color=\smallencoderc, mark=square*, mark size=3pt}, 
    spladedocplot/.style={thick, color=\noencoderc, mark=*, mark size=3pt}, 
    spladedocstoppedplot/.style={thick, color=\noencoderc, mark=diamond*, mark size=3pt},    bmvinplot/.style={thick, color=\bmvinc, mark=x, only marks, mark size=3pt}, 
    bmvinstoplot/.style={thick, color=\bmvinc, mark=triangle*, only marks, mark size=3pt}, 
    docTplot/.style={thick, color=\docTc, mark=x, only marks, mark size=3pt}, 
    deepimpacplot/.style={thick, color=\deepimpacc, mark=star, only marks, mark size=3pt}, 
    unicoilplot/.style={thick, color=\unicoilc, mark=*, only marks, mark size=3pt},
    flatplot/.style={thick, color=\docTc, mark=triangle*, only marks, mark size=3pt}, 
    ivfplot/.style={thick, color=\baselineplotc, mark=triangle*, mark size=3pt},
    hnswplot/.style={thick, color=black, mark=triangle*,mark size=3pt},
    dotivfplot/.style={dashed, color=\baselineplotc, mark=triangle, mark size=3pt},
    dothnswplot/.style={dashed, color=black, mark=triangle,mark size=3pt}}

\usepackage{pgfkeys}
\usepackage[normalem]{ulem}
\useunder{\uline}{\ul}{}

\newenvironment{customlegend}[1][]{%
    \begingroup
    \csname pgfplots@init@cleared@structures\endcsname
    \pgfplotsset{#1}%
}{%
    \csname pgfplots@createlegend\endcsname
    \endgroup
}%
\def\addlegendimage{\csname pgfplots@addlegendimage\endcsname}

\copyrightyear{2022}
\acmYear{2022}
\setcopyright{acmlicensed}\acmConference[SIGIR '22]{Proceedings of the 45th
International ACM SIGIR Conference on Research and Development in
Information Retrieval}{July 11--15, 2022}{Madrid, Spain}
\acmBooktitle{Proceedings of the 45th International ACM SIGIR Conference on
Research and Development in Information Retrieval (SIGIR '22), July 11--15,
2022, Madrid, Spain}
\acmPrice{15.00}
\acmDOI{10.1145/3477495.3531833}
\acmISBN{978-1-4503-8732-3/22/07}

\begin{document}
\title{An Efficiency Study for SPLADE Models}


 \author{Carlos Lassance}
 \affiliation{%
   \institution{Naver Labs Europe}
   \city{Meylan}
   \country{France}
  }

 \email{first.last at naverlabs.com}

 \author{Stéphane Clinchant}
 \affiliation{%
   \institution{Naver Labs Europe}
   \city{Meylan}
   \country{France}}
 \email{first.last at naverlabs.com}






\renewcommand{\shortauthors}{Lassance et Clinchant}

\begin{abstract}
Latency and efficiency issues are often overlooked when evaluating IR models based on Pretrained Language Models (PLMs) in reason of multiple hardware and software testing scenarios. Nevertheless, efficiency is an important part of such systems and should not be overlooked.

In this paper, we focus on improving the efficiency of the SPLADE model since it has achieved state-of-the-art zero-shot performance and competitive results on TREC collections.  SPLADE efficiency can be controlled via a regularization factor, but solely controlling this regularization has been shown to not be efficient enough. In order to reduce the latency gap between SPLADE and traditional retrieval systems, we propose several techniques including L1 regularization for queries, a separation of document/query encoders, a FLOPS-regularized middle-training, and the use of faster query encoders. Our benchmark demonstrates that we can drastically improve the efficiency of these models while increasing the performance metrics on in-domain data. To our knowledge, {we propose the first neural models that, under the same computing constraints, \textit{achieve similar latency (less than 4ms difference) as traditional BM25}, while having \textit{similar performance (less than 10\% MRR@10 reduction)} as the state-of-the-art single-stage neural rankers on in-domain data}.

\end{abstract}


\settopmatter{printacmref=true} 

\maketitle

\section{Introduction}
As search engines process billion of queries every day, efficiency has been a long standing research topic in Information Retrieval (IR). For instance, optimizing an inverted index with compression techniques or adopting an efficient two stage ranking pipeline to improve performance while maintaining an acceptable latency are commonplace for most search engines. Today, the advances of Pretrained Language Models (PLMs) have challenged the foundations of many ranking systems, which are based on term-based approaches like BM25 for first-stage ranking \cite{robertson2009probabilistic}.  Among this new generation of first-stage rankers, there exist two types of models. On one hand, the \textit{dense} retrieval models \cite{xiong2021approximate,qu-etal-2021-rocketqa,Hofstaetter2021_tasb_dense_retrieval,lin-etal-2021-batch} rely on Approximate Nearest Neighbors (ANN) techniques developed by the Computer Vision community.
On the other hand, the \textit{sparse} retrieval models \cite{dai2019contextaware,nogueira2019document,zhao2020sparta,sparterm2020,gao-etal-2021-coil,10.1145/3404835.3463030,10.1145/3404835.3463098} perform matching at the token level and therefore use a traditional inverted index for scoring.
 
 If TREC competitions and other retrieval benchmarks report performance measures such as NDCG@10 and have shown the immense benefit of pretrained language models, the overall picture for latency is less clear.
 In fact, measuring the latency of these novel models is challenging as there could be multiple testing conditions. For instance, a standard dense bi-encoder may rely on multiple CPUs to perform the search while some systems rely only on a single core and others on multi-core implementations.
 An  advantage of sparse retrieval models is the vast literature~\cite{ding2011faster, mallia2017faster, dimopoulos2013optimizing, khattab2020finding, mallia2021fast,turtle1995query,broder2003efficient,lin2015anytime} in optimizing retrieval with inverted indices. Furthermore, these works achieve impressive mono-cpu retrieval numbers for traditional sparse retrieval models~\cite{mackenzie2020efficiency}, making it simple to improve the scalability of the system (one just needs to add more cpus). This differs vastly from the context of dense retrieval, where multi-threaded (and sometimes even GPU~\cite{Hofstaetter2021_tasb_dense_retrieval}) seems to be the norm. Second, integrating a sparse ranker into an existing system may be less costly compared to the integration of a dense retrieval system. Note that even if the dense systems use a lot more compute (multi-cpu core + GPU vs mono-cpu core), the average latency\footnote{Considering the MSMARCO dataset of 8.8M passages and the systems used in the respective papers} from exact search of dense models on gpu (e.g TAS-B has retrieval+inference of 64 ms for 34.3 MRR@10)~\cite{Hofstaetter2021_tasb_dense_retrieval} tend to be equivalent than the one of sparse models on mono-cpu (e.g unicoil-T5 has retrieval + inference 36 ms + 45 = 81ms for 35.2 MRR@10)~\cite{wacky}. We include a short comparison on the appendix, but will focus on sparse retrieval for the rest of this work.

In this work, we focus on the SPLADE model\footnote{We include in the appendix a discussion on how these improvements may be included in other sparse models} as it was shown to be a state-of-the-art sparse retrieval model~\cite{splade_v2_arxiv}. Regarding latency, the model was actually optimized for multi-thread retrieval and no numbers were given, until a recent study \cite{wacky} reported that SPLADE was not well suited to mono-cpu retrieval environment. In this paper we aim at reducing this gap in performance, and thus focus on reducing mono-thread retrieval of SPLADE models under the PISA~\cite{mallia2019pisa} and ANSERINI~\cite{yang2018anserini} frameworks. Our main contribution is the proposal of 4 adaptations to the SPLADE model, namely: \begin{inlinelist}\item separation of query and document encoders \item change the query regularization to L1 \item a FLOPS-regularized PLM \item the use of a smaller PLM for query encoding \end{inlinelist}. In order to showcase the utility of our adaptations, we focus on three main research questions:
\begin{itemize}
    \item RQ1: How does the performance (efficiency and effectiveness) of SPLADE evolve as we introduce each subsequent adaptation?
    \item RQ2: How do these new SPLADE models compare to the state of the art in terms of in-domain sparse retrieval?
    \item RQ3: How do these new SPLADE models compare to the state of the art in terms of out-of-domain sparse retrieval?
\end{itemize}

\section{Previous Works}
\textbf{Measuring efficiency of PLM-based  retrieval systems:} Comparing efficiency metrics between methods is a complicated task. While benchmarking effectiveness is easy because all methods focus on solving the same task, and thus we can find ``one'' metric that is comparable between all methods, evaluating efficiency is naturally a trade-off (how much impact does it have on the effectiveness metric). Efficiency depends on which type of system it is focused on (multi-thread or mono-thread per query retrieval) and also depends heavily on the machine used to perform the measures~\cite{hofstatter2019let}. For example, efficient sparse retrieval is a domain in itself, with a rich diversity of methods~\cite{ding2011faster, mallia2017faster, dimopoulos2013optimizing, khattab2020finding, mallia2021fast,turtle1995query,broder2003efficient,lin2015anytime} that have been proposed in order to improve their retrieval times. Note that the ``best method'' depends on various factors~\cite{wacky,khattab2020finding} and there is no one method that is better than all the others. In this work, our focus is to propose adaptations to the SPLADE model and not to the retrieval process itself, thus we avoid these open questions by using the document-at-a-time retrieval setup from previous works~\cite{mackenzie2020efficiency,wacky} in order to perform our study. However, we believe that jointly improving the PLM-based model and the retrieval process should lead to even more improved performance.

\textbf{PLM-based dense retrieval:} Recently, IR tasks have been taken by an avalanche of PLM-based dense retrievers~\cite{xiong2021approximate,qu-etal-2021-rocketqa,Hofstaetter2021_tasb_dense_retrieval,lin-etal-2021-batch,gao2021unsupervised}. However, due to their dense nature, efficiency studies have been mostly confined on multi-cpu or even GPU-based systems~\cite{Hofstaetter2021_tasb_dense_retrieval} and will therefore not be considered here as they are not comparable.  For instance, \cite{Hofstaetter2021_tasb_dense_retrieval} reports ``quick'' methods that have less than 70ms in multi-cpu/GPU systems, while we consider models with less than 10ms on mono-cpu settings. Furthermore, when we consider dense retrieval models, they either have increased latency/space constraints (for example ColBERTv2~\cite{santhanam2021colbertv2}, even with the more efficient PLAID approach~\cite{santhanam2022plaid}) or have been shown to have lower performances on out-of-domain retrieval~\cite{beir_2021}. Note that there seems to be a new solution for the latter, which calls for a large-batch contrastive pre/middle-training of the models, which seems to aliviate (but not completely correct) this problem as seen in Contriever~\cite{izacard2021contriever} and LaPraDoR~\cite{xu2022laprador}.

\textbf{PLM-based sparse retrieval:} Another way to perform retrieval using PLMs is to use a lexical (sparse) base~\cite{dai2019contextaware,nogueira2019document,zhao2020sparta,sparterm2020,gao-etal-2021-coil,10.1145/3404835.3463030,10.1145/3404835.3463098,splade_v2_arxiv}. These systems take advantage of the PLMs to perform document and (sometimes) query expansion while also doing term-reweighting. Among the previously cited sparse models, SPLADE~\cite{splade_v2_arxiv} has shown the best performance in out-of-domain tasks.
However, it was originally presented with the same multi-cpu techniques than the dense models, which  leads actually to large latencies on mono-cpu conditions ~\cite{wacky}. We thus use SPLADE as the basis for this work, where we aim to improve its efficiency on mono-cpu retrieval. In the following, we do a quick summary on SPLADE.

\textbf{SPLADE:}
It uses the BERT~\cite{bert} token space to predict term importance ($|V|\approx 30k$). These term importances are based on the logits of the Masked Language Model (MLM).
Let $d$ a document and $q$ a query, and let $w_{ij}$ be the logit for $i$th token in $d$
for the probability of term $j$.
In other words, weight $w_{ij}$ is how important the PLM considered the term $j$ of the token space to the input token $i$.
SPLADE then takes the importance for each token in the document sequence and max pools them, to generate a  vector in the BERT vocabulary space.
SPLADE models are then optimized via distillation. These models jointly optimize: \begin{inlinelist} \item distance between teacher and student scores
\item minimize the expected mean FLOPS of the retrieval system \end{inlinelist}. This joint optimization can be described as: 
\begin{equation}
\mathcal{L} = \mathcal{L}_{distillation} + \lambda_q \mathcal{L}^{q}_{\texttt{FLOPS}} + \lambda_d \mathcal{L}^{d}_{\texttt{FLOPS}}
\end{equation} 
where $\mathcal{L}_{\texttt{FLOPS}}$ is the sparse \texttt{FLOPS} regularization from~\cite{paria2020minimizing} and $\mathcal{L}_{distillation}$ is a distillation loss between the scores of a teacher and a student (in this work we use KL Divergence~\cite{hinton2015distilling,kullback1951information} as the loss and a cross-ranker~\cite{passage_ranking} as teacher). Note that there are two distinct regularization weights, so that one can put more sparsity pressure in either queries or documents, but always considering the amount of FLOPS. 

\paragraph{\bf SPLADE-doc:}
In~\cite{splade_v2_arxiv} the authors also propose to consider a document-only version of SPLADE, i.e without query encoder. In this case, there is no query expansion, 
only tokenization is done, with all query terms having the same importance.


\section{Methods}

Analyzing the setup from the multi-cpu retrieval used in SPLADEv2~\cite{splade_v2_arxiv} and the mono-cpu in~\cite{wacky}, we derive that the main source of improvement is the reduction of SPLADE query sizes\footnote{amount of tokens at the SPLADE output}, instead of focusing solely on the FLOPS measure. The reason is that in mono-threaded systems, there are many techniques that allow for reducing the amount of effective FLOPS computed per query, but query size is then a major bottleneck. To get faster and more effective SPLADE models, we propose the following improvements: \begin{inlinelist}[label=\roman*)]
    \item Searching for appropriate hyperparameters
    \item Improving data used for training
    \item Separating document and query encoders
    \item Changing query regularization to L1
    \item Middle training of a PLM with a FLOPS regularization
    \item Smaller PLM query encoder
\end{inlinelist}.

Of those adaptations, i) and ii) are just used in order to find better baselines.
On the other hand, iii), iv), v) and vi) are novel contributions. Note that, we consider \textit{each adaptation sequentially}, so that for example vi) is the system with all the proposed adaptations.

\subsection{Baselines}

\paragraph{\bf{i) Searching for appropriate hyperparameters:}}
First of all,  we checked if we could get more efficient networks with SPLADE as we change some training hyperparameters. Notably, we change the distillation loss from MarginMSE~\cite{hofstatter2020improving} to the more classical KL Divergence~\cite{kullback1951information,hinton2015distilling} loss as in our studies it was more stable. Then, we also search for a set of $(\lambda_q, \lambda_d)$ in order to have controlled query and document sizes. At the end of that experiment, we chose to keep a set of three configuration: \textbf{S}mall, \textbf{M}edium, \textbf{L}arge, where (S) is the sparsest configuration, (M) has the same query sparsity as S, but larger documents and (L) has larger queries than S and the same document sparsity as M.

\paragraph{\bf{ii) Improving data used for training:}}
The second improvement comes from changing the data and model used for distillation. The goal is to improve the effectiveness of the networks chosen on the previous experiment, while avoiding increasing the cost of inference. To do so, we move from the more traditional set of distillation data from~\cite{hofstatter2020improving} to a newer one available from hugginface \footnote{ in~\url{https://huggingface.co/datasets/sentence-transformers/msmarco-hard-negatives}}. The main difference is while the first one uses negatives from BM25, the latter uses negatives coming from various sources (dense models) and a more powerful teacher (trained with distillation).

\subsection{Efficient SPLADE Models}
With better baselines for SPLADE, let us introduce some changes to the overall model in order to improve its efficiency:

\paragraph{\bf{iii) Separating document and query encoders:}}

While searching for proper baselines, we noticed that it was hard to achieve smaller queries, as even very large differences in $\lambda_d$ and $\lambda_q$ did not produce smaller queries. This is because the encoder for both documents and queries is the same and there is nothing to differentiate between them. Therefore we propose to have two distinct encoders for queries and documents, thus relieving the model to find an optimal trade-off for document and queries with a single model.

\paragraph{\bf{iv)} L1 Regularization for Queries:}
A second change is to reconsider the FLOPS measure. While it make sense for document representation, it may not be the best measure accounting for latency of retrieval system. 
This is why we propose to change the regularization type of our query encoder to a simple L1 loss rather than the FLOPs on the query vectors. 

\paragraph{\bf{v) PLM Middle Training  with FLOPS regularization:}}

Recently there has been a surge~\cite{gao2021unsupervised,izacard2021towards} in what we call ``middle-training'' of PLMs for IR. The idea is that before fine-tuning on a dataset, one should refine a pre-trained model on the data and task that it will perform, with the most known example being Co-Condenser~\cite{gao2021unsupervised} for dense networks, that performs two steps of middle training: i) condensing information into the CLS with MLM ii) training jointly an unsupervised contrastive and MLM losses~\cite{bert}. In this work, we investigate if performing a middle-training step, where we use an MLM task with a FLOPS regularization on the MLM logits, improves the final fine-tuned SPLADE model.

\paragraph{\bf{vi) Smaller PLM query encoder:}}

An important factor when computing the latency of PLM-based models is the latency of the query encoder. This could be indeed very costly considering that we are on the mono-threaded, no-GPU configuration. In the SPLADEv2 paper~\cite{splade_v2_arxiv} the authors propose to use a scheme without any query encoder (called SPLADE-doc). In this work, we evaluate two methods: \begin{inlinelist} \item remove the stop words of the queries and retrain SPLADE-doc model to further improve this method (we note by $^\dagger$ the systems that remove the stop words of queries) \item use a very efficient PLM on the query encoder, namely BERT-tiny~\cite{bhargava2021generalization}~\footnote{Another possibility would be to perform quantization and/or compression of the PLM, but we leave exploring this to future work}. \end{inlinelist}

\section{Experimental setting and results}

We trained and evaluated our models on the MS MARCO passage ranking dataset~\cite{Bajaj2016Msmarco} in the full ranking setting. The dataset contains approximately $8.8$M passages, and hundreds of thousands training queries with shallow annotation ($\approx 1.1$ relevant passages per query in average). The development set contains $6980$ queries with similar labels. 
We also consider the full BEIR benchmark~\cite{beir_2021} (18 datasets) which judges the zero-shot performance of IR models over diverse set of tasks and domains.

\paragraph{\bf{Measuring efficiency}:}

In order to compute our efficiency numbers, all experiments are performed in the same machine, with an Intel(R) Xeon(R) Gold 6338 CPU @ 2.00GHz CPU and sufficient RAM memory to preload indexes, models and queries on memory before starting the experiment. Experiments are performed using Anserini and PISA\footnote{In the main paper we report PISA and leave Anserini for the appendix.} for retrieval (following instructions obtained by contacting the authors of~\cite{wacky}) and PyTorch for document/query encoding. All efficiency experiments with PyTorch use the benchmarking tool from the transformers library~\cite{wolf-etal-2020-transformers}. 

The latency of query encoder PLMs was measured using a sequence length of 8 for average latency and 32 for 99 percentile latency. Latency is computed as a simple addition of query encoding and retrieval time. The DistilBERT query encoder has an average latency of $45.3$ms and a 99 percentile latency of $57.6$ms, while the BERT-tiny~\cite{bhargava2021generalization} query encoder has an average latency of $0.7$ms and a 99 percentile latency of $1.1$ms. 


\paragraph{\bf SPLADE Training:}

We use \texttt{DistilBERT-base} as the starting point for most models, safe for the improvements \emph{v}) and \emph{vi}), which use a middle-trained DistilBERT and middle-trained BERT-tiny. SPLADE Models are trained for 250k steps with the ADAM optimizer, using a learning rate of $2e^{-5}$ with linear scheduling and a warmup of $6000$ steps, and a batch size of $128$. We keep the last step as our final checkpoint. For the SPLADE-doc approach, we follow~\cite{splade_v2_arxiv} and perform a reduced training of only $50$k steps. We consider a maximum length of $256$ for input sequences. In order to mitigate the contribution of the regularizer at the early stages of training, we follow \cite{paria2020minimizing} and use a scheduler for $\lambda$, quadratically increasing $\lambda$ at each training iteration, until a given step ($50$k for SPLADE and $10$k for SPLADE-doc), from which it remains constant. Middle-training is performed using default MLM parameters from~\cite{wolf2020huggingfaces}, with an added FLOPS regularization~\cite{paria2020minimizing} of $\lambda=0.001$. Concerning ($\lambda_q,\lambda_d$), models i), ii), iii) use the same hyperparameters: S$=(0.1,5e-3)$, M$=(0.1,5e-4)$, and L$=(0.01,5e-4)$, while models iv), v) and vi) use S=$(5e-3,5e-3)$, M$=(5e-4,5e-4)$, L$=(5e-4,5e-4)$.

\paragraph{\bf{RQ1: How does the performance (efficiency and effectiveness) of SPLADE evolve as we introduce each subsequent adaptation?}}

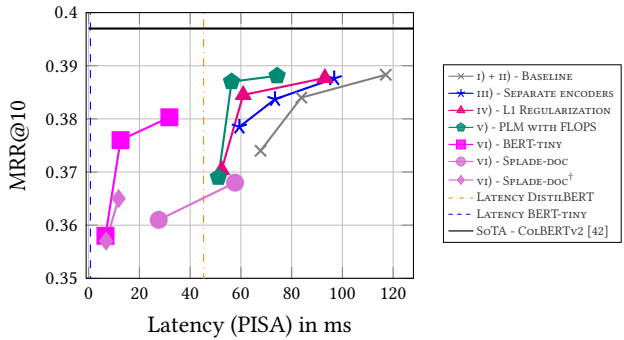
\begin{figure}[ht]
     \centering
     \begin{subfigure}[t]{0.7\columnwidth}

         \centering
\adjustbox{max width=\textwidth}{%
            \begin{tikzpicture}
       \begin{axis}[
           xlabel=Latency (PISA) in ms,
           ylabel=MRR@10,
           xmin=-1, xmax=128, ymin=0.35, ymax=0.40]
         \addplot[baselineplot] table {figures/rq1/pisa_results_2.txt};
         \addplot[separateplot] table {figures/rq1/pisa_results_3.txt};
         \addplot[lplotc] table {figures/rq1/pisa_results_4.txt};
         \addplot[plmplotc] table {figures/rq1/pisa_results_5.txt};
         \addplot[tinybertplot] table {figures/rq1/pisa_results_6.txt};
         \addplot[spladedocplot] table {figures/rq1/pisa_splade_doc.txt};
         \addplot[spladedocstoppedplot] table {figures/rq1/pisa_splade_doc_stopped.txt};
        \draw[dash dot, orange] (45.3,0) -- (45.3,1);
        \draw[dashed, blue] (0.7,0) -- (0.7,1);
        \draw[thick, black] (0,0.397) -- (1000,0.397);
       \end{axis}
    \end{tikzpicture}
         }
     \end{subfigure}
     \begin{subfigure}[t]{0.29\columnwidth}
         \centering
\adjustbox{max width=\textwidth}{%
         \raisebox{70px}{%
         \begin{tikzpicture}
\begin{customlegend}[
legend entries={\textsc{i) + ii) - Baseline} ,
                        \textsc{iii) - Separate encoders} ,
                        \textsc{iv) - L1 Regularization},
                        \textsc{v) - PLM with FLOPS},
                        \textsc{vi) - BERT-tiny},
                        \textsc{vi) - Splade-doc},
                        \textsc{vi) - Splade-doc$^\dagger$},
                        \textsc{Latency DistilBERT},
                        \textsc{Latency BERT-tiny},
                         \textsc{SoTA - ColBERTv2~\cite{santhanam2021colbertv2}},
                        }]
        \addlegendimage{baselineplot}
        \addlegendimage{separateplot}   
        \addlegendimage{lplotc}
        \addlegendimage{plmplotc}   
        \addlegendimage{tinybertplot}
        \addlegendimage{spladedocplot}   
        \addlegendimage{spladedocstoppedplot}
        \addlegendimage{dash dot, orange}   
        \addlegendimage{dashed, blue}   
        \addlegendimage{thick, black}
        \end{customlegend}
\end{tikzpicture}}
}
     \end{subfigure}
      \caption{Latency comparison between all proposed improvements. $^\dagger$: queries without stop words.}
     \label{fig:rq1}
\end{figure}

First we verify the relevance of adding each subsequent adaptation. Note that each ``level'' [i), ii), iii)...] \textbf{represents its change and all the others that came before}. We represent the efficiency and effectiveness of the systems in Figure~\ref{fig:rq1}, using pytorch for PLM inference and PISA for mono-cpu retrieval. For each model we have three points, representing each combination of $\lambda_d,\lambda_q$ that we introduced in the previous section. SPLADEv2-distil is not shown on the Figure because it made it hard to read, with a reported latency of 265~\cite{wacky} and a total latency measured on our system of 691 ms.

First, we see that just using a stronger baseline allows us to decrease the latency in PISA by almost 10 times (same in Anserini) of the latency of SPLADEv2-distil, while keeping similar or even improved performance on MS MARCO. This is expected, because the model used for comparison in~\cite{wacky} was not optimized for latency. We see also that all models trained are close to the single-stage state-of-the-art retrieval performance of ColBERTv2~\cite{santhanam2021colbertv2}(at most a 10\% reduction in MRR@10 performance).

Second, we see that improvements iii) through v) each successfully improve latency, while keeping similar effectiveness on MS MARCO. Note however, that each one seems to bring diminishing gains, because they are heavily impacted by the inference latency of DistilBERT, especially on PISA, for example, sparsest v) has a gain of 1.2ms compared to sparsest iv), which represents an approximate 20\% reduction retrieval time (PISA), but an overall reduction of just 2\% (PISA+Pytorch) . In order to mitigate this, we introduce vi) which aims at speeding up the query encoder.

With vi) we note a trade-off, that at the cost of a slight effectiveness loss ($\approx$1.0 MRR@10 on MS MARCO), we are able to greatly reduce the latency of the sparsest SPLADE models we test ($\approx$10 fold PISA, $\approx$2 fold Anserini). For the query-encoder choice, BERT-tiny had a slight advantage over SPLADE-doc, showing the importance of a query-encoder, even if it is a very small one.

\paragraph{\bf{RQ2: How do these new SPLADE models compare to the state of the art in terms of in-domain sparse retrieval?}}

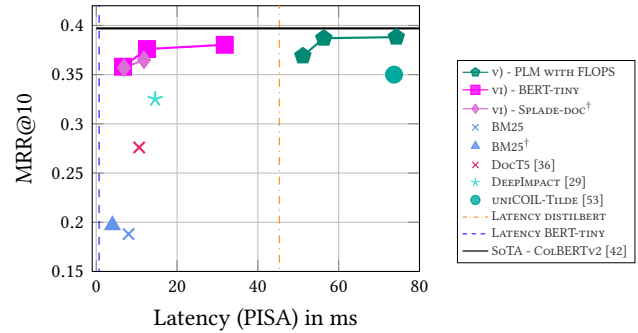
\begin{figure}[ht]
     \centering
     \begin{subfigure}[t]{0.7\columnwidth}

         \centering
\adjustbox{max width=\textwidth}{%
            \begin{tikzpicture}
       \begin{axis}[
           xlabel=Latency (PISA) in ms,
           ylabel=MRR@10,
           xmin=-1, xmax=80, ymin=0.15, ymax=0.42,
           ]
         \addplot[plmplotc] table {figures/rq1/pisa_results_5.txt};
         \addplot[tinybertplot] table {figures/rq1/pisa_results_6.txt};
         \addplot[spladedocstoppedplot] table {figures/rq1/pisa_splade_doc_stopped.txt};
         \addplot[bmvinplot] table {figures/rq2/pisa_bm25.txt};
         \addplot[bmvinstoplot] table {figures/rq2/pisa_bm25_stopped.txt};
         \addplot[docTplot] table {figures/rq2/pisa_docT5.txt};
         \addplot[deepimpacplot] table {figures/rq2/pisa_deepimpact.txt};
         \addplot[unicoilplot] table {figures/rq2/pisa_unicoil-tilde.txt};
         
        \draw[dash dot, orange] (45.3,0) -- (45.3,1);
        \draw[dashed, blue] (0.7,0) -- (0.7,1);
        \draw[thick, black] (0,0.397) -- (1000,0.397);
       \end{axis}
    \end{tikzpicture}
         }
     \end{subfigure}
     \begin{subfigure}[t]{0.29\columnwidth}
         \centering
\adjustbox{max width=\textwidth}{%
         \raisebox{50px}{%
         \begin{tikzpicture}
\begin{customlegend}[
legend entries={\textsc{v) - PLM with FLOPS},
                        \textsc{vi) - BERT-tiny},
                        \textsc{vi) - Splade-doc$^\dagger$},
                        \textsc{BM25},
                        \textsc{BM25$^\dagger$},
                        \textsc{DocT5~\cite{doct5}},
                        \textsc{DeepImpact~\cite{10.1145/3404835.3463030}},
                        \textsc{uniCOIL-Tilde~\cite{zhuang2021fast}},
                        \textsc{Latency distilbert},
                         \textsc{Latency BERT-tiny},
                         \textsc{SoTA - ColBERTv2~\cite{santhanam2021colbertv2}},
                        }]
        \addlegendimage{plmplotc}   
        \addlegendimage{tinybertplot}
        \addlegendimage{spladedocstoppedplot}
         \addlegendimage{bmvinplot};
         \addlegendimage{bmvinstoplot};
         \addlegendimage{docTplot};
         \addlegendimage{deepimpacplot};
         \addlegendimage{unicoilplot};
        \addlegendimage{dash dot, orange}   
        \addlegendimage{dashed, blue}   
        \addlegendimage{thick, black}
        \end{customlegend}
\end{tikzpicture}}
}
     \end{subfigure}
      \caption{Latency comparison between the proposed adaptations and sparse methods. $^\dagger$: queries without stop words.}
     \label{fig:rq2}
\end{figure}

In the previous question, we verified that the proposed improvements made sense in the context of SPLADE, but what does that mean against other methods in sparse retrieval? To answer this question we compare a subset of our systems \{v) and vi)\} to the baselines used in~\cite{wacky}, namely: \begin{inlinelist} \item BM25 \item DocT5~\cite{doct5} \item DeepImpact~\cite{10.1145/3404835.3463030} \item uniCOIL-Tilde~\cite{zhuang2021fast} \end{inlinelist}. All methods are evaluated on our machine and DistilBERT latency is added to uniCOIL-Tilde. 
Results are displayed in Figure~\ref{fig:rq2}.

We can see that compared to the non-BM25 techniques we can achieve systems that are both more efficient and have better effectiveness for in-domain sparse retrieval\footnote{Note that some of the improvements proposed here could be applied to these systems, check the appendix for a discussion}.  Finally, compared to BM25 we achieve similar efficiency, with a 2x gain on effectiveness. 

\paragraph{\bf{RQ3: How do these new SPLADE models compare to the state of the art in terms of out-of-domain sparse retrieval?}}

In the previous sections we showed the efficiency and effectiveness of the proposed adaptations of SPLADE in-domain, i.e. on MS MARCO. However, one of the main advantages of the SPLADEv2-distil model was in out-of-domain retrieval, notably on the BEIR benchmark (at the time of submission SoTA on the online benchmark). We now study the effect of our improvements, with MS MARCO MRR@10 and BEIR mean nDCG@10 results shown in Table~\ref{table:rq3}\footnote{A detailed per dataset result table is available in the Appendix.}.

\begin{table}[ht]
\centering
\caption{Results on out-of-domain data and MS MARCO efficiency. Results for MSMARCO are given as MRR@10 on the 6980 dev set queries and nDCG@10 over the TREC-19~\cite{craswell2020overview} queries, while BEIR results are the average nDCG@10. BEIR* is the combination of BM25 and the row's method. BT: BERT-tiny query encoder. $^\dagger$: queries without stop-words.  $^\mathsection$: Differs from~\cite{splade_v2_arxiv} as we consider the 18 beir datasets.}
\label{table:rq3}
\adjustbox{max width=\columnwidth}{%
\begin{tabular}{r|c|cc|cc}
\toprule
\multicolumn{1}{c|}{Method} &                     Latency & MSMARCO & TREC19 & BEIR & BEIR$^\star$ \\ \midrule
\multicolumn{5}{c}{Baselines}                                                                                                                                                                                                                                                                                                               \\ \midrule
BM25$^\dagger$                                  & \textbf{4}                                        & 19.7    & 50.6 & 43.0  & -   \\
DocT5~\cite{doct5}                                     & 11                    & 27.6  & 64.2 & 44.1 & -  \\
SPLADEv2-distil~\cite{splade_v2_arxiv}                                  & 691  & 36.8                        & \textbf{72.9} & \textbf{47.0}$^\mathsection$ & \textbf{49.3}    \\ \midrule
\multicolumn{5}{c}{Proposed models}                                                                                                                                                                                                                             \\ \midrule
VI) BT-SPLADE-S                    & {\ul \textit{7}} & 35.8 & 67.2 & 39.2 & 45.9  \\
VI) BT-SPLADE-M                   & 13 & {\ul \textit{37.6}} & 69.4 & 42.1 & 47.1  \\
VI) BT-SPLADE-L                & 32 & \textbf{38.0} & {\ul \textit{70.3}}   & {\ul \textit{44.5}} & {\ul \textit{48.0}}  \\ 

 \bottomrule
\end{tabular}}
\end{table}

Unfortunately, the gains in efficiency and in-domain effectiveness seem to come at a cost of reducing the performance of the models outside of the MS MARCO domain. While it is still has decent effectiveness compared to BM25 (which is not the case for most dense models), it still loses a lot of performance when compared to SPLADEv2-distil. Investigating the performances per dataset we noticed some sharp losses on some datasets, especially on the QUORA dataset (where nDCG@10 falls from 0.84 on SPLADEv2-distil to 0.46 on BT-SPLADE-S), which is expected as it uses queries (questions) as both documents \emph{and} queries. 

One way to mitigate this overall loss of effectiveness is by merging the document scores of the proposed models with the ones obtained by BM25. This comes at a cost, either adding the latency of BM25 (4ms) or by duplicating the computing cost but keeping the latency of the slowest model. We use a simple score combination inspired by~\cite{lin-etal-2021-batch}, where documents not present in the top-100 of a model are assigned its smallest score and then normalizing the scores based on the maximum and minimum document scores of the method. Finally we simply sum the scores of the two methods, assigning equal weight to both. We represent the ensemble as column BEIR* on the table. We see that it allows us to slightly outperform SPLADEv2-distil by itself, while running under 40 ms of latency on a single cpu-core. On the more efficient models, combining our method with BM25 allows us to outperform DocT5 on BEIR on similar latency (11ms). 

Finally, we also note that SPLADEv2-distil combined with BM25 is able to outperform methods that came after it~\cite{neelakantan2022text,muennighoff2022sgpt,xu2022laprador} and had claimed the ``state-of-the-art''. In other words, it is, as far as we are aware, the best result available on BEIR (49.3 mean on all 18 datasets).

\section{Conclusion}

In this paper we investigated the efficiency of SPLADE models and proposed small adaptations that led to more efficient systems. We demonstrated that the proposed adaptations improved both the efficiency and effectiveness of SPLADE for in-domain performance and that the small degradatation  on out-of-domain performance can be mitigated by combining with traditional sparse retrieval techniques. We also note that some of these adaptations could be applied to other systems, but that comparison is left as future work. In summary, to our knowledge, {we propose the first neural models that \textit{achieve similar mono-cpu latency and multi-cpu throughput as traditional BM25}, while having \textit{similar performance} than state-of-the-art first-stage neural rankers on in-domain data (MS MARCO) and comparable performance on the (BEIR) benchmark to both BM25 and to most dense first-stage neural rankers}.
\begin{acks}
We would like to thank Thibault Formal, Hervé Dejean, Jean-Michel Renders and Simon Lupart for helping with their thoughtfull comments. We also would like to thank the authors of Wacky Weights~\cite{wacky} (Joel Mackenzie, Andrew Trotman and Jimmy Lin) for sharing their PISA evaluation code with us.
\end{acks}

\bibliographystyle{ACM-Reference-Format}
\bibliography{sample-base}

\appendix

\section{On the use of mono-cpu average latency as a metric and exclusion of dense comparisons}

Throughout the paper we used mono-cpu average latency as the defacto-metric and excluded dense models from the comparison. We do benchmarking in this way to make comparisons as fair as possible, as it allows us to control as many variables as possible. In the following we give our reasoning for those decisions, which are up for discussion and are not necessarily the best ones\footnote{Note to the reader: I would be glad to discuss this further carlos.lassance@naverlabs.com}.

In this work we follow~\cite{wacky} for the benchmarking protocol. In their work they also use mono-cpu average latency as the main metric for efficiency and this decision seems reasonable to us. For a search system there will always be a balance of amount of computation available vs amount of incoming requests, and focusing in ``atomic'' performance allows for easier scalability later (increasing QPS is easy by just increasing the amount of computation available). It also makes it simple to compare with numbers previously reported in the literature. In the following we will also report QPS using 128 cpus (i.e. we treat 128 queries in parallel) and show the relative power of this implementation in a more realistic scenario. However, even those two metrics still obfuscate possible problems with long tail queries and index size/maximum concurrent memory requirements. 

Moreover, including dense models would only increase the amount of comparison problems. Notably the trade-offs are different (less of a problem with long tail queries, increased index size/memory requirements) and an implementation problem arises (can we say that X representation is faster or is just an advantage of Y implementation being faster?). We develop more on this types of problems in the next section.

\section{Comparison against dense models}

In the main paper we compare mostly against sparse models and only report the effectiveness of a SoTA dense multi-representation model (ColBERTv2~\cite{santhanam2021colbertv2}). We do so because we do not feel there is a proper way to compare both methods (for example all our latency numbers use PISA for retrieval and Pytorch for inference, which is impossible in the case of dense models). For completeness in this section we investigate a comparison between a SoTA mono-representation dense model (CoCondenser~\cite{gao2021unsupervised} available at \url{https://huggingface.co/Luyu/co-condenser-marco-retriever}) and the models studied in this work. 

In order to study different settings that make different trade-offs we define three: \begin{inlinelist} \item Measuring latency under mono-cpu using one query at a time \item Measuring QPS (queries per second) under multi-threaded-cpu using batched queries \item Measuring QPS (queries per second) multi-threaded-cpu disregarding inference time with batched queries \end{inlinelist}. The dense model uses the FAISS~\cite{johnson2019billion} framework, where we either do brute-force or perform ANN using HNSW~\cite{malkov2018efficient} and IVF~\cite{sivic2003video} indexes, without any quantization\footnote{We tested with product quantization and the efficiency-effectiveness trade-off was worse than without. As we do not show index size on the figures we preferred to only deal with non quantized models in order to reduce the number of parameters to tune}. While we don't feel that any of the three settings are fully fair to both approaches (they fail to acknowledge index size, maximum RAM and ANN use for sparse retrieval~\cite{wacky, tu2020approximate}), we feel that overall they represent some of the different scenarios that one would consider for deploying such systems. 

\begin{figure}[ht]
     \centering
     \begin{subfigure}[t]{0.7\columnwidth}

         \centering
\adjustbox{max width=\textwidth}{%
            \begin{tikzpicture}
       \begin{axis}[
           xlabel=QPS (PISA/FAISS),
           ylabel=MRR@10,
           xmin=0, ymin=0.35, ymax=0.4,    xmode=log,
                      log ticks with fixed point,log basis x={4},
                      xmin=4,xmax=8200, xtick={8,32,128,512,2048,8192},
                      xticklabels = {8,32,128,512,2048,8192}
           ]
         \addplot[plmplotc] table {figures/qps/with_inference/splade_5.txt};
         \addplot[tinybertplot] table {figures/qps/with_inference/splade_6.txt};
         \addplot[hnswplot] table {figures/qps/with_inference/hnsw32_bert.txt};
         \addplot[ivfplot] table {figures/qps/with_inference/ivf65536_bert.txt};         
         \draw[thick, \docTc] (0,0.383) -- (10000,0.383);
       \end{axis}
    \end{tikzpicture}
         }
     \end{subfigure}
     \begin{subfigure}[t]{0.29\columnwidth}
         \centering
\adjustbox{max width=\textwidth}{%
         \raisebox{100px}{%
         \begin{tikzpicture}
\begin{customlegend}[
legend entries={\textsc{v) - PLM with FLOPS},
                        \textsc{vi) - BERT-tiny},
                        \textsc{Brute Force - CC},
                        \textsc{HNSW32 - CC},
                        \textsc{IVF65536 - CC},
                        }]
        \addlegendimage{plmplotc}   
        \addlegendimage{tinybertplot}
        \addlegendimage{thick, \docTc}
         \addlegendimage{hnswplot};
         \addlegendimage{ivfplot};
         \addlegendimage{dothnswplot};
         \addlegendimage{dotivfplot};
        \end{customlegend}
\end{tikzpicture}}
}
     \end{subfigure}
      \caption{QPS comparison between the proposed adaptations and dense methods with ANN, considering inference time. Results are better up and to the right. xaxis in $log_4$ scale. DB-CoCondenser omitted as difference with CoCondenser is too small. CC: CoCodenser }
     \label{fig:qps_with_inference}
\end{figure}
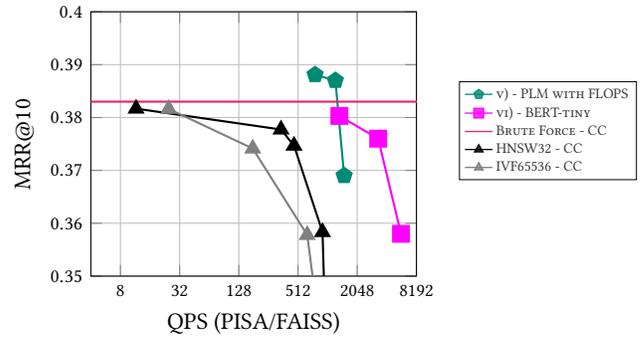

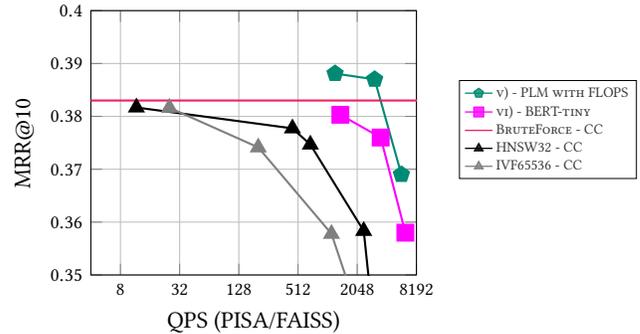
\begin{figure}[ht]
     \centering
     \begin{subfigure}[t]{0.7\columnwidth}

         \centering
\adjustbox{max width=\textwidth}{%
            \begin{tikzpicture}
       \begin{axis}[
           xlabel=QPS (PISA/FAISS),
           ylabel=MRR@10,
           xmin=0, ymin=0.35, ymax=0.4,    xmode=log,
                      log ticks with fixed point,log basis x={4},
                      xmin=4,xmax=8200, xtick={8,32,128,512,2048,8192},
                      xticklabels = {8,32,128,512,2048,8192}
           ]
         \addplot[plmplotc] table {figures/qps/without_inference/splade_5.txt};
         \addplot[tinybertplot] table {figures/qps/without_inference/splade_6.txt};
         \draw[thick, \docTc] (0,0.383) -- (10000,0.383);
         \addplot[hnswplot] table {figures/qps/without_inference/hnsw32.txt};
         \addplot[ivfplot] table {figures/qps/without_inference/ivf65536.txt};
         
       \end{axis}
    \end{tikzpicture}
         }
     \end{subfigure}
     \begin{subfigure}[t]{0.29\columnwidth}
         \centering
\adjustbox{max width=\textwidth}{%
         \raisebox{100px}{%
         \begin{tikzpicture}
\begin{customlegend}[
legend entries={\textsc{v) - PLM with FLOPS},
                        \textsc{vi) - BERT-tiny},
                        \textsc{BruteForce - CC},
                        \textsc{HNSW32 - CC},
                        \textsc{IVF65536 - CC},
                        }]
        \addlegendimage{plmplotc}   
        \addlegendimage{tinybertplot}
        \addlegendimage{thick, \docTc}
         \addlegendimage{hnswplot};
         \addlegendimage{ivfplot};
        \end{customlegend}
\end{tikzpicture}}
}
     \end{subfigure}
      \caption{QPS comparison between the proposed adaptations and dense methods with ANN, disregarding inference time. Results are better up and to the right. xaxis in $log_4$ scale. CC: CoCondenser.}
     \label{fig:qps_without_inference}
\end{figure}

Also note that the dense model already starts at a disavantage due to the fact that it uses a larger encoder (BERT, around two times the computational cost of distilBERT) and that as far as we are aware there are no SoTA dense implementations using BERT-tiny. In order to consider a distilBERT version of CoCondenser we assume a \emph{perfect} quantization of the BERT model into distilBERT and a scenario where inference is disregarded, which are not realistic, but give an idea of the best possible numbers are if we had "perfect" inference. Results are depicted in Figures~\ref{fig:dense_latency},~\ref{fig:qps_with_inference}, and~\ref{fig:qps_without_inference}.

\begin{figure}[ht]
     \centering
     \begin{subfigure}[t]{0.7\columnwidth}

         \centering
\adjustbox{max width=\textwidth}{%
            \begin{tikzpicture}
       \begin{axis}[
           xlabel=Latency in ms (PISA/FAISS),
           ylabel=MRR@10,
           xmin=4, ymin=0.35, ymax=0.4, xmode=log, log ticks with fixed point,        log basis x={2}, xtick={4,8,16,32,64,128,256,512}
           ]
         \addplot[plmplotc] table {figures/rq1/pisa_results_5.txt};
         \addplot[tinybertplot] table {figures/rq1/pisa_results_6.txt};
         \draw[thick, \docTc] (0,0.383) -- (1000,0.383);
         \addplot[dothnswplot] table {figures/qps/latency/hnsw32.txt};
         \addplot[dotivfplot] table {figures/qps/latency/ivf65536.txt};
         \addplot[hnswplot] table {figures/qps/latency/hnsw32_bert.txt};
         \addplot[ivfplot] table {figures/qps/latency/ivf65536_bert.txt};

       \end{axis}
    \end{tikzpicture}
         }
     \end{subfigure}
     \begin{subfigure}[t]{0.29\columnwidth}
         \centering
\adjustbox{max width=\textwidth}{%
         \raisebox{100px}{%
         \begin{tikzpicture}
\begin{customlegend}[
legend entries={\textsc{v) - PLM with FLOPS},
                        \textsc{vi) - BERT-tiny},
                        \textsc{BruteForce - CC},
                        \textsc{HNSW32 - CC},
                        \textsc{IVF65536 - CC},
                        \textsc{HNSW32 - DB-CC},
                        \textsc{IVF65536 - DB-CC},
                        }]
        \addlegendimage{plmplotc}   
        \addlegendimage{tinybertplot}
        \addlegendimage{thick, \docTc}
         \addlegendimage{hnswplot};
         \addlegendimage{ivfplot};
         \addlegendimage{dothnswplot};
         \addlegendimage{dotivfplot};
        \end{customlegend}
\end{tikzpicture}}
}
     \end{subfigure}
      \caption{Latency comparison between the proposed adaptations and dense methods with ANN. Results are better up and to the left. xaxis in $log_2$ scale. CC: CoCondenser. DB CoCondenser is an ideal quantization of CoCondenser from BERT to distilBERT.}
     \label{fig:dense_latency}
\end{figure}
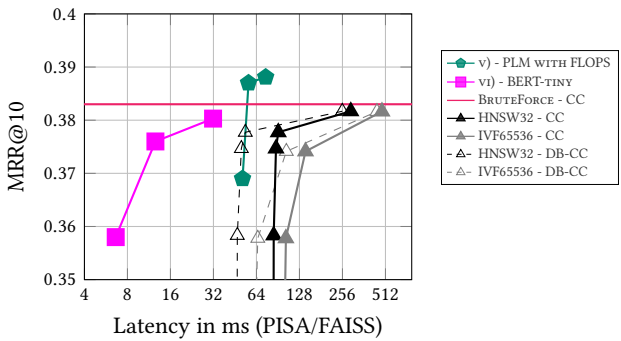

\begin{table*}[ht]
\caption{Detailed BEIR results using 100 times the nDCG@10 for each dataset. BT: BERT-tiny query encoder and BT-{S/M/L} is the VI) BT-SPLADE-{S/M/L} model. $^\dagger$: model uses queries without stop-words. Bold represents best in row.}
\label{beir-detail}
\adjustbox{max width=\textwidth}{
\begin{tabular}{c||ccc|cc||ccc|ccc}
\toprule
\multirow{2}{*}{Dataset} & \multicolumn{5}{c||}{Baselines}                                    & \multicolumn{6}{c}{Proposed}                                                                                                   \\ 
                         & BM25$^\dagger$  & BM25          & DocT5 & SPLADEv2-distil       & SPLADEv2-distil+BM25$^\dagger$  & BT-S & BT-M & BT-L & BT-S + BM25$^\dagger$ & BT-M + BM25$^\dagger$ & BT-L + BM25$^\dagger$ \\ \midrule
arguana                  & 42.25 & 41.42          & 46.90 & 47.91          & 48.71          & 45.47           & 46.58           & 47.29           & 47.28                  & 49.16                  & \textbf{50.12}         \\
bioasq                   & 47.67 & 46.46          & 43.10 & 50.80          & \textbf{55.22} & 41.17           & 45.20           & 47.13           & 50.69                  & 53.09                  & 53.72                  \\
climate-fever            & 21.32 & 21.29          & 20.10 & 23.53          & \textbf{26.59} & 17.09           & 18.00           & 18.89           & 22.81                  & 23.34                  & 23.58                  \\
cqadupstack              & 28.53 & 29.87          & 32.50 & \textbf{35.01} & 34.49          & 28.22           & 30.52           & 33.01           & 31.35                  & 32.53                  & 33.70                  \\
dbpedia-entity           & 32.26 & 31.28          & 33.10 & \textbf{43.50} & 41.01          & 35.38           & 38.91           & 40.54           & 37.99                  & 39.35                  & 40.09                  \\
fever                    & 74.35 & 75.31          & 71.40 & 78.62          & \textbf{82.44} & 69.88           & 72.14           & 74.87           & 79.13                  & 79.29                  & 80.04                  \\
fiqa                     & 24.30 & 23.61          & 29.10 & \textbf{33.61} & 32.45          & 28.78           & 30.07           & 31.77           & 30.89                  & 31.15                  & 31.80                  \\
hotpotqa                 & 60.13 & 60.28          & 58.00 & 68.44          & \textbf{70.02} & 56.59           & 61.31           & 66.61           & 63.80                  & 66.34                  & 68.44                  \\
nfcorpus                 & 32.67 & 32.55          & 32.80 & 33.43          & \textbf{34.07} & 31.52           & 32.43           & 33.10           & 33.16                  & 33.50                  & 33.70                  \\
nq                       & 32.87 & 32.86          & 39.90 & \textbf{52.08} & 48.20          & 47.35           & 49.69           & 51.48           & 46.25                  & 47.24                  & 48.33                  \\
quora                    & 74.71 & 78.86          & 80.20 & 83.76          & \textbf{84.99} & 46.29           & 62.24           & 72.34           & 68.45                  & 74.86                  & 78.54                  \\
robust04                 & 41.91 & 40.84          & 43.70 & 46.75          & \textbf{49.52} & 33.73           & 37.07           & 40.66           & 44.02                  & 45.64                  & 46.60                  \\
scidocs                  & 15.83 & 15.81          & 16.20 & 15.79          & \textbf{16.92} & 14.56           & 14.56           & 15.25           & 16.43                  & 16.47                  & 16.80                  \\
scifact                  & 66.28 & 66.47          & 67.50 & 69.25          & \textbf{71.58} & 63.94           & 65.77           & 67.43           & 68.66                  & 69.59                  & 70.45                  \\
signal1m                 & 32.69 & 33.05          & 30.70 & 26.56          & 33.31          & 24.50           & 27.21           & 28.28           & 31.56                  & 33.47                  & \textbf{33.51}         \\
trec-covid               & 71.23 & 65.59          & 71.30 & 71.04          & \textbf{76.89} & 59.91           & 64.38           & 66.06           & 73.67                  & 75.64                  & 75.63                  \\
trec-news                & 40.33 & 39.77          & 42.00 & 39.18          & \textbf{45.46} & 35.43           & 34.80           & 38.65           & 43.30                  & 43.03                  & 43.96                  \\
webis-touche2020         & 35.40 & \textbf{36.73} & 34.70 & 27.18          & 35.85          & 26.54           & 26.42           & 27.03           & 33.53                  & 34.89                  & 35.39                  \\
\midrule
Average                     & 43.04 & 42.89          & 44.07 & 47.02          & \textbf{49.32} & 39.24           & 42.07           & 44.47           & 45.85                  & 47.14                  & 48.02                  \\
Best on                  & 0     & 1              & 0     & 4              & \textbf{11}    & 0               & 0               & 0               & 0                      & 0                      & 2           \\          
\bottomrule
\end{tabular}
}
\end{table*}

In mono-cpu latency the brute force dense approach takes too long compared to all others (more than 200 ms), so we do not depict it on the image. We thus have to compare our sparse retrieval methods with ANN-based dense retrieval\footnote{We could also apply ANN to sparse retrieval~\cite{tu2020approximate,wacky}, but we leave this as future work}. We were actually surprised that HNSW allows dense models to be very close to the efficient SPLADE models (if we consider perfect quantization from BERT to distilBERT), however, HNSW comes with a drawback of increased index size: in our case 256 bytes extra per passage or around 2GB for MSMARCO\footnote{The entire SPLADE pisa index takes around 1GB for S, 2GB for M and 4GB for L}. If we consider methods that do not increase index size by much (such as IVF) the difference is still pretty large\footnote{Unquantized dense indexes are around 25GB for fp32 and 12GB for fp16}. Under multiple cpu threads we still see an advantage for the models presented in this work, even when compared to BERT-based CoCodenser (Fig~\ref{fig:qps_with_inference}) or even if we disregard inference time (Fig~\ref{fig:qps_without_inference}. These results further show the interest of studying sparse representations (equal/better efficiency-effectiveness trade-off with smaller index).

\section{Detailed BEIR results}

For completeness, we now present the BEIR results, dataset per dataset, in Table~\ref{beir-detail}. We can see that SPLADEv2-distil actually dominates all other approaches in the amount of datasets that it is the best on (4 by itself and 11 alongside BM25) and that BM25 by itself is only the best in webis-touche2020. The combination of VI)BT-SPLADE-M and L with BM25 present more balanced results, having less datasets that they outperform the others compared with SPLADEv2-distil, however they have a similar/better average effectiveness for less than 5\% the cost of running SPLADEv2-distil.

\section{Latency results using Anserini}

\begin{figure}[ht]
     \centering
     \begin{subfigure}[t]{0.7\columnwidth}

         \centering
\adjustbox{max width=\textwidth}{%
            \begin{tikzpicture}
       \begin{axis}[
           xlabel=Latency (Anserini) in ms,
           ylabel=MRR@10,
           xmin=-10, xmax=600, ymin=0.35, ymax=0.40]
         \addplot[baselineplot] table {figures/rq1/anserini_results_2.txt};
         \addplot[separateplot] table {figures/rq1/anserini_results_3.txt};
         \addplot[lplotc] table {figures/rq1/anserini_results_4.txt};
         \addplot[plmplotc] table {figures/rq1/anserini_results_5.txt};
         \addplot[tinybertplot] table {figures/rq1/anserini_results_6.txt};
         \addplot[spladedocplot] table {figures/rq1/anserini_splade_doc.txt};
         \addplot[spladedocstoppedplot] table {figures/rq1/anserini_splade_doc_stopped.txt};
        \draw[dash dot, orange] (45.3,0) -- (45.3,1);
        \draw[dashed, blue] (0.7,0) -- (0.7,1);
        \draw[thick, black] (0,0.397) -- (1000,0.397);
       \end{axis}
    \end{tikzpicture}
         }
     \end{subfigure}
     \begin{subfigure}[t]{0.29\columnwidth}
         \centering
\adjustbox{max width=\textwidth}{%
         \raisebox{70px}{%
         \begin{tikzpicture}
\begin{customlegend}[
legend entries={\textsc{i) + ii) - Baseline} ,
                        \textsc{iii) - Separate encoders} ,
                        \textsc{iv) - L1 Regularization},
                        \textsc{v) - PLM with FLOPS},
                        \textsc{vi) - BERT-tiny},
                        \textsc{vi) - Splade-doc},
                        \textsc{vi) - Splade-doc$^\dagger$},
                        \textsc{Latency DistilBERT},
                        \textsc{Latency BERT-tiny},
                         \textsc{SoTA - ColBERTv2~\cite{santhanam2021colbertv2}},
                        }]
        \addlegendimage{baselineplot}
        \addlegendimage{separateplot}   
        \addlegendimage{lplotc}
        \addlegendimage{plmplotc}   
        \addlegendimage{tinybertplot}
        \addlegendimage{spladedocplot}   
        \addlegendimage{spladedocstoppedplot}
        \addlegendimage{dash dot, orange}   
        \addlegendimage{dashed, blue}   
        \addlegendimage{thick, black}
        \end{customlegend}
\end{tikzpicture}}
}
     \end{subfigure}
      \caption{Latency comparison between all proposed improvements. $^\dagger$: queries without stop words.}
     \label{fig:rq1_anserini}
\end{figure}
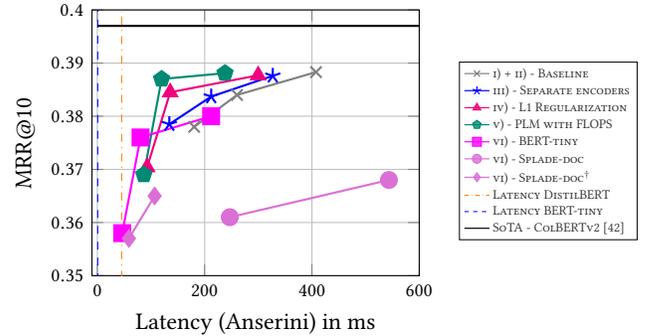

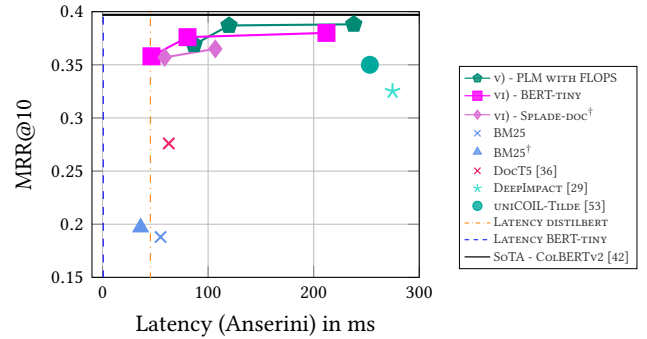
\begin{figure}[ht]
     \centering
     \begin{subfigure}[t]{0.7\columnwidth}

         \centering
\adjustbox{max width=\textwidth}{%
            \begin{tikzpicture}
       \begin{axis}[
           xlabel=Latency (Anserini) in ms,
           ylabel=MRR@10,
           xmin=-10, xmax=300, ymin=0.15, ymax=0.40,
           ]
         \addplot[plmplotc] table {figures/rq1/anserini_results_5.txt};
         \addplot[tinybertplot] table {figures/rq1/anserini_results_6.txt};
         \addplot[spladedocstoppedplot] table {figures/rq1/anserini_splade_doc_stopped.txt};

         \addplot[bmvinplot] table {figures/rq2/anserini_bm25.txt};
         \addplot[bmvinstoplot] table {figures/rq2/anserini_bm25_stopped.txt};
         \addplot[docTplot] table {figures/rq2/anserini_docT5.txt};
         \addplot[deepimpacplot] table {figures/rq2/anserini_deepimpact.txt};
         \addplot[unicoilplot] table {figures/rq2/anserini_unicoil-tilde.txt};

        \draw[dash dot, orange] (45.3,0) -- (45.3,1);
        \draw[dashed, blue] (0.7,0) -- (0.7,1);
        \draw[thick, black] (0,0.397) -- (1000,0.397);
       \end{axis}
    \end{tikzpicture}
         }
     \end{subfigure}
     \begin{subfigure}[t]{0.29\columnwidth}
         \centering
\adjustbox{max width=\textwidth}{%
         \raisebox{50px}{%
         \begin{tikzpicture}
\begin{customlegend}[
legend entries={\textsc{v) - PLM with FLOPS},
                        \textsc{vi) - BERT-tiny},
                        \textsc{vi) - Splade-doc$^\dagger$},
                        \textsc{BM25},
                        \textsc{BM25$^\dagger$},
                        \textsc{DocT5~\cite{doct5}},
                        \textsc{DeepImpact~\cite{10.1145/3404835.3463030}},
                        \textsc{uniCOIL-Tilde~\cite{zhuang2021fast}},
                        \textsc{Latency distilbert},
                         \textsc{Latency BERT-tiny},
                         \textsc{SoTA - ColBERTv2~\cite{santhanam2021colbertv2}},
                        }]
        \addlegendimage{plmplotc}   
        \addlegendimage{tinybertplot}
        \addlegendimage{spladedocstoppedplot}
         \addlegendimage{bmvinplot};
         \addlegendimage{bmvinstoplot};
         \addlegendimage{docTplot};
         \addlegendimage{deepimpacplot};
         \addlegendimage{unicoilplot};
        \addlegendimage{dash dot, orange}   
        \addlegendimage{dashed, blue}   
        \addlegendimage{thick, black}
        \end{customlegend}
\end{tikzpicture}}
}
     \end{subfigure}
      \caption{Latency comparison between the proposed adaptations and sparse methods. $^\dagger$: queries without stop words.}
     \label{fig:rq2_anserini}
\end{figure}

In the main paper we presented results using PISA, which can be seen as a best case scenario for sparse representations. A more real life scenario is to consider Anserini, which is based on Lucene and is more production ready~\cite{}. In Figure~\ref{fig:rq1_anserini} and Figure~\ref{fig:rq2_anserini} we show the effectiveness and efficiency of each sucessive improvement and then compare with the other state of the art sparse representation solutions. However, Anserini brings forward two new questions 1) the latency of DistilBERT is not as much of a drawback, making the distance between VI) and V) models less important; 2) If we re-compare with dense models the conclusions would be different, which was one of the reasons we preferred to avoid making direct comparisons with dense in the first place, as the conclusion depends on the implementation of the retrieval system (the same could happen if instead of using FAISS~\cite{faiss} we used other implementations for dense retrieval).

\section{Improvements on other sparse models}

In this paper we perform benchmarking against other sparse models as they are found in the literature. This leads to a comparison that is not necessarily fair, as the same improvements that we make for SPLADE could be applied for them. We note that our main objective was to compare these improvements on the SPLADE model itself and then re-position it compared to BM25, which cannot make use of these improvements.

While such a benchmarking methodology is a common place in the literature (distillation was initially only applied to dense models~\cite{hofstatter2020improving}, query clustering for better in-batch negatives has only been applied to one specific model~\cite{Hofstaetter2021_tasb_dense_retrieval} and many other examples), we would like to at least acknowledge the other models and discuss where these modifications could apply:

\subsection{DeepImpact}

DeepImpact~\cite{10.1145/3404835.3463030} is a method for generating sparse representations using the docT5~\cite{doct5} expansion as a base. There is no query encoder and words are stored in the same way as traditional indexes, i.e. BERT tokenization is not applied. Considering that information, and the fact that the model is not trained with distillation, improvement i), ii) could be easily applied to this model (changing the hyperparameters/initial networks and using distillation/better training data). Using better PLM models (Implictly in V) could also help DeepImpact, but we are not sure what is the best way (Contriever~\cite{izacard2021contriever}, CoCondenser~\cite{gao2021unsupervised}, MLM+Flops?). Most of the improvements do not apply as there's no sparsity regularization (expansion comes from docT5) and no query encoder. 

\subsection{uniCOIL}

uniCOIL~\cite{zhuang2021fast} is another method that generates sparse representations, using either docT5~\cite{doct5} or TILDEv2~\cite{zhuang2021fast} to generate the document expansion, controlled by a fixed parameter. Note that query expansion is not performed, but words are still stored using BERT tokenization, which means that it has to be applied during query inference. As it is the case for DeepImpact, uniCOIL could benefit from improvements i) and ii). As it is the case for DeepImpact, better PLMs should help, but a better study would have to be done to define in which way. Finally the other improvements do not apply, as it does not have a query encoder and does not use sparsity regularization.

\end{document}